\documentclass[%
 reprint,
 amsmath,amssymb,
 aps,
]{revtex4-2}

\usepackage{graphicx}% Include figure files
\usepackage{dcolumn}% Align table columns on decimal point
\usepackage{bm}% bold math
\usepackage{tikz}
\usepackage{hyperref}

\begin{document}

\preprint{APS/123-QED}

\title{Using molecular dynamics to investigate the driving force for graphene auto-kirigami}% Force line breaks with \\

\author{Charlie M. Rawlins}
\author{Gareth A. Tribello}%
 \email{g.tribello@qub.ac.uk}
\affiliation{%
 Centre for Quantum Materials and Technologies, School of Mathematics and Physics, Queen's University Belfast, Belfast, BT7 1NN, United Kingdom
}%

\date{\today}% It is always \today, today,

\begin{abstract}
Experiments \cite{Cross_16} have shown that auto-kirigami structures can grow on the surface of graphene because the graphene-graphene adhesion energy is greater than the graphene-substrate interaction. In this work molecular dynamics (MD) simulations of folded graphene both in vacuum and on a substrate have been performed for a range of different initial geometries and at various temperatures. The final equilibrated configuration for many of these simulations resembles a book with a fold that is at the midpoint of the graphene sheet. We investigate the amount of time it takes to move from an initial folded configuration that does not have the fold at the midpoint of the graphene to this final, book-like configuration.  We show that graphene-graphene and graphene substrate-adhesion energies can be extracted from such simulations and that the values obtained are almost always higher than the values of these quantities that are obtained from static calculations. We also show that the rate at which these folded structures grow is affected by differences in these adhesion energies. However, this rate is also affected by factors such as the initial geometry of the graphene that do not change the adhesion energies. Understanding the kinetics of auto-kirigami formation thus requires a description of the system that is more sophisticated than an energy-balance model. 
%\begin{description}
%\end{description}
\end{abstract}

\maketitle
\section{Introduction\label{sec:intro}}

%MAIN RESULT: ENERGY IN FOLDING MODEL CANNOT BE CALCULATED FROM STATIC CALCULATIONS OF GRAPHENE-GRAPHENE AND GRAPHENE SUBSTRATE INTERACTIONS.

%NEW TITLE: USING MOLECULAR DYNAMICS TO INVESTIGATE THE DRIVING FORCE FOR GRAPHENE AUTO-KIRIGAMI 

Graphene's discovery has sparked extensive research activity across physics, chemistry, and materials science\cite{Yanwu_10}. As graphene is one atom thick, it exhibits remarkable properties that stem directly from its ultra-thin structure\cite{Neto_09}. Of particular interest are the exceptional mechanical flexibility within the plane of the sheet and the strong van der Waals forces between layers which enable graphene to be manipulated into complex configurations such as scrolls \cite{Viculis_03, Pereira_21} and folds\cite{Zhang_10}. These graphene assemblies demonstrate fascinating characteristics, with superlubricity being among the most compelling phenomena observed\cite{Dienwiebel_lubricity_04,Yilun_14,Popov_11_commensurate,Popov_11_commensurateb}.

\begin{figure}
    \centering
    \includegraphics[width=\linewidth]{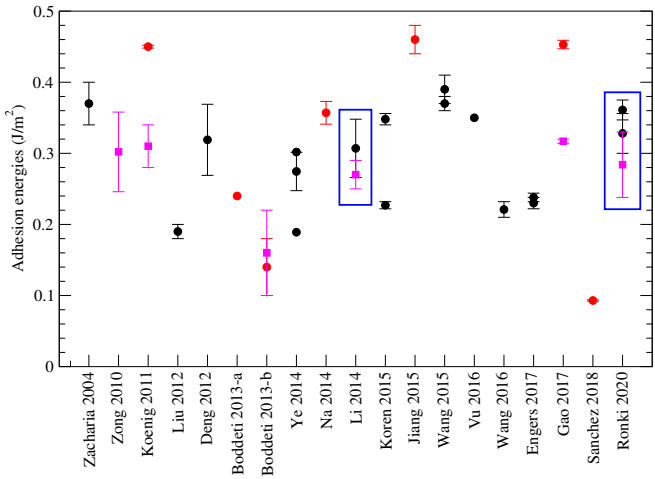}
    \caption{The black points in this figure show the experimental measurements for the graphene-graphene adhesion energy, $\gamma_{gg}$ \cite{Zacharia_04,Liu_12_deformation,Deng_12,Ye_2014,Li_2014,Koren2015adhesion,Wang_2015,Wang_16,Vu_2016,Engers_17,Rokni_2020}  The red and purple points show single-layer and bilayer graphene-substrate adhesion energies, $\gamma_{gs}$, respectively\cite{zong2010direct,Koenig_2011,Boddeti_2013a,Boddeti_2013b,Na_14,Li_2014,Jiang_2015,Gao_2017,Sanchez_18,Rokni_2020}.  The squares around certain data points highlight the studies that measured both $\gamma_{gg}$ and $\gamma_{gs}$. }
    \label{fig:exp}
\end{figure}

A particularly interesting graphene phenomenon has recently been discovered by Annett and Cross who have developed a method for synthesizing folded graphene\cite{Cross_16}. They first disrupt the graphene structure using a nanoindentor or AFM tip.  Novel kirigami ribbons then grow spontaneously from the disrupted region. In the continuum mechanics models that they have used to rationalize the geometry of the structures that form, the graphene-graphene adhesion energy, $\gamma_{gg}$, is larger than the graphene-substrate energy, $\gamma_{gs}$.  In fact, this difference in energy is large enough to compensate for the energy, $\lambda$, required to break the carbon-carbon bonds that occurs during tearing. This result appears remarkable but as Annet and Cross argue \cite{Cross_16} the energy of formation for a rectangular pleat with a folding energy of $U_{fold}$ and a length of $c$ and a width of $w$ is given by:

\begin{equation}
E = U_{fold} -\gamma_{gs}wl_x+ ( \gamma_{gs} - \gamma_{gg} )w c + 2\lambda c.
\label{eqn:driving}
\end{equation}

Differentiating this expression with respect to $c$ and setting $\gamma_{gg}=0.361$ Jm$^{-2}$, $\gamma_{gs}=0.284$ Jm$^{-2}$, \cite{Rokni_2020} and $\lambda=4.8 \times 10^{-9}$ Jm$^{-1}$ \cite{Cross_16} we find that $\frac{\partial E}{\partial c}$ is negative once the pleat has a width that is greater than 125 nm. In other words, if we use the most recent experimental measurements for $\gamma_{gg}$ and $\gamma_{gs}$ the pleat grows as the energy that is gained by increasing amount of graphene-graphene contact grows much more rapidly than the energy lost through bond breaking.  

When making the calculations in the last paragraph we used the most recent estimates of the graphene-graphene and graphene-substrate adhesion energies. Figure \ref{fig:exp} shows the various estimates for these quantities that have been obtained from experiments. You can see that a range of values for the adhesion energies have been obtained.  Furthermore, when the data from all these studies is aggregated in figure \ref{fig:exp} we see results that suggest graphene binds more tightly to the substrate than it does to itself. If this were the case the model introduced above suggests that there is no driving force for autokirigami formation. Importantly, however, the majority of studies measured either $\gamma_{gg}$ or $\gamma_{gs}$. The two studies  \cite{Li_2014,Rokni_2020} that measured both found that $\gamma_{gg}>\gamma_{gs}$ and thus showed that the formation of kirigami features is energetically favorable. 

Using the simple energy balance model outlined in the previous paragraph to understand kirigami pleat formation obviously neglects many of the microscale phenomena that might impact upon pleat formation.  Consequently, in this work we have used molecular dynamics (MD) to test whether or not models like the one described above reproduce what is observed in direct simulations of graphene folding.  As discussed in section \ref{subsec:nofold} static calculations suggest that $\gamma_{gg}>\gamma_{gs}$ but this difference is very small for our model. Even so, we show that folded structures grow in our simulations in section \ref{sec:twist} and \ref{sec:temperature}. In section \ref{sec:extract} we therefore expand on the models from our previous paper on graphene in vacuum \cite{Rawlins_24} and show how values for $\gamma_{gg}$ and $\gamma_{gs}$ can be obtained from direct simulations of fold formation. The values for $\gamma_{gg}$ that we get from this analysis are almost always higher than the values that we get from static simulations. However, these $\gamma_{gg}$ values, like those from static calculations, are not affected by the initial geometry of the graphene or the temperature.

\section{Adhesion energies}
\label{subsec:nofold}

We ran MD simulations in the NVT ensemble at 300~K of single layer and bilayer graphene in vacuum and on an amorphous Si substrate using the Large-scale Atomic/Molecular Massively Parallel Simulator (LAMMPS\cite{LAMMPS}). The graphene sheets we simulated measured 600 by 300 \AA, while the Si substrate measured 900, 400 and 15 \AA\ in the $x$, $y$ and $z$ directions respectively. The AIREBO \cite{AIREBO_2000} potential was used to model carbon-carbon interactions and a Stillinger and Weber potential\cite{SW_85} was used for silicon silicon interactions. A Lennard-Jones potential with $\epsilon=9.6$meV and $\sigma=3.0$\AA\ \cite{Suzhi_16} was used to model the carbon silicon interaction.  Ensemble averages for the energy were computed from 50~ps simulations of these various systems that used a timestep of 0.125 fs and a Nose-Hoover thermostat\cite{Thermostat} with relaxation time of 12.5 fs.  We found that the energy equilibrated after 10 ps of simulation we thus used 40 ps of the simulation for our production calculations.

\begin{figure}
    \centering
    \includegraphics[width=\linewidth]{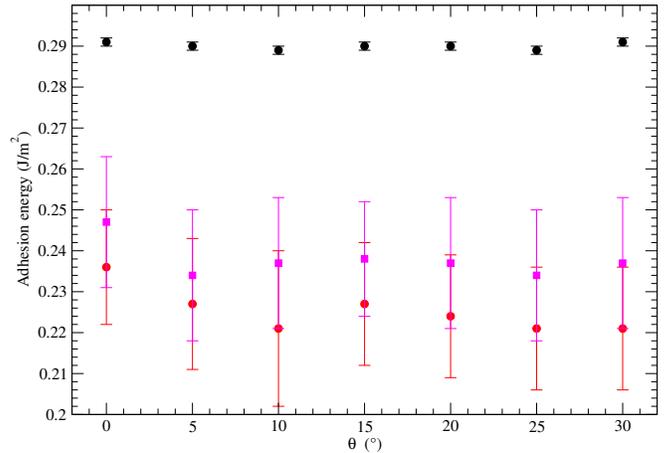}
    \caption{Calculated values for $\gamma_{gg}$ (black), $\gamma_{gs}^{SL}$ (red) and $\gamma_{gs}^{BL}$ (magenta) for the potential used in this work as a function of twist angle, $\theta$. $\gamma_{gg}$ values are calculated from simulations of bilayer graphene in vacuum.  Changing the twist angle for these structures involves rotating the top and bottom layers of the graphene through $-\theta$ and $\theta$ respectively. The same procedure is used when preparing the simulations of bilayer graphene on the substrate.  Obviously, only one graphene sheet is rotated when we perform the simulations of single layer graphene on the substrate.}
    \label{fig:layeradhesion}
\end{figure}

Average values for the energy of single layer graphene in vacuum, $E_{SL}$, and on a substrate, $E_{SL}^{sub}$, bilayer graphene in vacuum, $E_{BL}$, and on a substrate, $E_{BL}^{sub}$ as well as the energy of the isolated substrate, $E^{sub}$, were then computed from energies that were stored every 0.25 ps.  The production part of these timeseries were divided into eight 5-ps blocks to get the errors on our estimates of the average energies.  To calculate the graphene-graphene, $\gamma_{gg}$, single-layer-graphene-substrate, $\gamma_{gs}^{SL}$ and bilayer-graphene substrate, $\gamma_{gs}^{BL}$ interaction energies we then used the following expressions:
\begin{eqnarray}
    \gamma_{gg}&=&-\frac{E_{BL}-2E_{SL}}{A} \\
    \gamma_{gs}^{SL} &=& -\frac{E_{SL}^{sub}-E^{sub}-E_{SL}}{A}\\
    \gamma_{gs}^{BL} &=& -\frac{E_{BL}^{sub}-E^{sub}-E_{BL}}{A}
\end{eqnarray}
where $A$ is the area of graphene sheet.

%\begin{table}[]
%    \centering
%    \begin{tabular}{c|c}
%         & Adhesion values (J/m$^2$) \\ \hline
%        $\gamma_{gg}$ &  0.291$\pm$0.001 \\
%        $\gamma_{gs}^{SL} $ & 0.236$\pm$0.014 \\
 %       $\gamma_{gs}^{BL}$ & 0.230$\pm$0.021
%    \end{tabular}
%    \caption{Adhesion values determined from static graphene layer simulations.}
%    \label{tab:layers}
%\end{table}

The twist angle ($\theta$) of the graphene is defined as the angle between the x-axis of the lab frame and the line of symmetry that, if cut, would give rise to an edge with an `arm-chair' configuration.  Adjusting this parameter is important because superlubricity is only observed for non-commensurately-stacked graphene sheets.  We thus calculated graphene-graphene and graphene-substrate adhesion energies for a range of different $\theta$ values. 
It is important to note that cutting graphene sheets along non arm-chair directions leads to structures in which some carbon atoms only participate in only one covalent bond.  In our simulations we removed these carbons and thereby ensured that all the carbon atoms at the edge of the graphene were bound to at least two additional carbon atoms.
The results are shown in figure \ref{fig:layeradhesion}, which clearly demonstrates that the adhesion energies do not depend on $\theta$.  This figure also  shows that the $\gamma_{gg}>\gamma_{gs}$ for the  potential we are using in this work. Equation \ref{eqn:driving} thus suggests that the formation of autokirigami pleats is favorable for this particular potential but that the energy difference that drives pleat formation is rather small.

\begin{figure}
{\begin{tikzpicture}
\draw (7,0) -- (1,0);
\draw[dashed] (7,0) -- (3,0) arc (270:90:0.25) -- (4.2,0.5);
\draw[|-|] (3,0.7) -- (4.2,0.7) node[anchor=west]{$\gamma_{gg}+\gamma_{gs}^{\rm BL}$};
\draw[|-|] (1,1.2) -- (4.2,1.2) node[anchor=west]{$-\gamma_{gs}^{\rm SL}$};
\node[draw=none] at (3.6,0.9) {$c$};
\node[draw=none] at (2.6,1.4) {$2c$};
\draw[ultra thick] (0,-0.5) -- (9,-0.5);
\end{tikzpicture}}
{\begin{tikzpicture}
\draw (7,0) -- (1,0) arc (270:90:0.25) -- (2.2,0.5);
\draw[dashed] (7,0) -- (3,0) arc (270:90:0.25) -- (6.2,0.5);
\draw[|-|] (4.2,1.2) -- (6.2,1.2)node[anchor=west]{$\gamma_{gg}+\gamma_{gs}^{\rm BL}$};
\draw[|-|] (2.2,1.7) -- (6.2,1.7)node[anchor=west]{$-\gamma_{gs}^{\rm SL}$};
\draw[ultra thick] (0,-0.5) -- (9,-0.5);
\node[draw=none] at (5.2,1.4) {$c$};
\node[draw=none] at (4.2,1.9) {$2c$};
\draw[|-|] (1,1.5) -- (2.2,1.5) node[midway,above] {$c_0$};
\draw[|-|] (1,0.7) -- (3,0.7) node[midway,above] {$c$};
\draw[|-|] (3,0.7) -- (4.2,0.7) node[midway,above] {$c_0$};
\end{tikzpicture}}\caption{Schematic image justifying equation \ref{eqn:new-driving}.  The top panel illustrates the change in conformation that occurs when a kirigami pleat with length $c$ nucleates. The energy change that accompanies this process, neglecting the contribution due to the bending of the graphene, is equal to $wc(2\gamma_{gs}^{SL} - \gamma_{gs}^{BL} - \gamma_{gg}$) as the process involves replacing a region of single layer that has an area of $2wc$ graphene with a piece of bilayer graphene with an area of $wc$. The bottom panel shows the conformational change that that occurs when a kirigami pleat grows by a length $c$. Increasing the area of bilayer graphene in the structure by $wc$ requires a $2wc$ decrease in the area of single layer graphene.}
\label{fig:motion}
\end{figure}
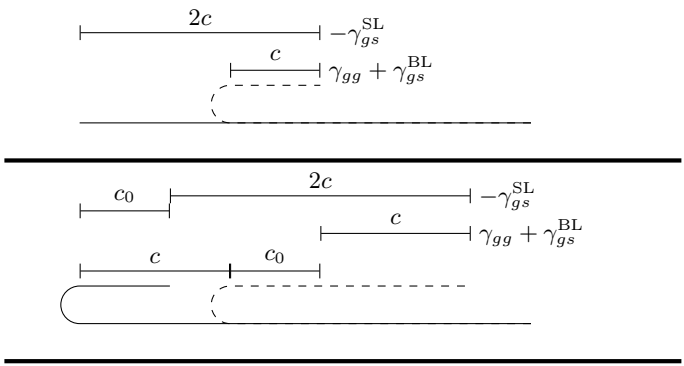

Equation \ref{eqn:driving} assumes that the top layer of any folded structure stops interacting with the underlying substrate.  To take account of interactions between the top layer of the autokirigami pleat and the underlying substrate, one could argue for rewriting equation \ref{eqn:driving} as:

\begin{equation}
E = U_{fold} -\gamma_{gs}wl_x + (2\gamma_{gs}^{SL} - \gamma_{gs}^{BL} - \gamma_{gg}) w c + 2\lambda c.
\label{eqn:new-driving}
\end{equation}

This expression gives the energy change that accompanies the initial process of fold formation that is illustrated in the top panel of figure \ref{fig:motion} and also suggests that fold growth is favorable for the model employed in this paper. The bottom panel of figure \ref{fig:motion} shows that this equation also holds if a fold has already formed even though it is \emph{bilayer} graphene that is peeling off from the substrate as it grows. The equation above is thus the appropriate way for calculating the energy change that accompanies the growth of these kirigami features.  

If we insert the line tension of $4.8 \times 10^{-9}$ Jm$^{-1}$ from \cite{Cross_16} and average adhesion energies from figure \ref{fig:layeradhesion} into equation \ref{eqn:new-driving} we find that autokirigami ribbons with a width of 196 nm should form with this potential. Running MD simulations of this size of graphene flake would be too computationally expensive.  Furthermore, as we showed in a previous paper \cite{Rawlins_24} the rate of growth of non-tearing, folded graphene structures does not depend on the width of the graphene sheet. Consequently, in the remainder of this paper we have simulated the growth of folded structures on graphene directly and determined the rate at which these folded structures grow.  As we will show, we find that graphene folds on substrates grow in agreement with what the static models would predict.

%A baseline test for calculating adhesion energy from simulations of graphene on a substrate would begin with flat sheets of graphene.
%For graphene-graphene adhension, the energy can be determined by running a simulation of bilayer (BL) graphene and of single layer (SL) graphene, taking the average energy of both runs, subtract the energy of two single layers from the energy of two layers to get the adhesion energy between the two. 
%Graphene-substrate adhesion works on a similar idea, run another simulation with the substrate by itself, then with a single layer of graphene on the substrate. Substrate from the total energy of the graphene on a substrate the energy of SL graphene by itself and the substrate by itself. We can also examine the difference with BL graphene. 
%The equations summarising this approach is shown below.

%% The values for the adhesion energies for these systems

\section{Methods}
\label{sec:method}
\begin{figure}
\includegraphics[width=\linewidth]{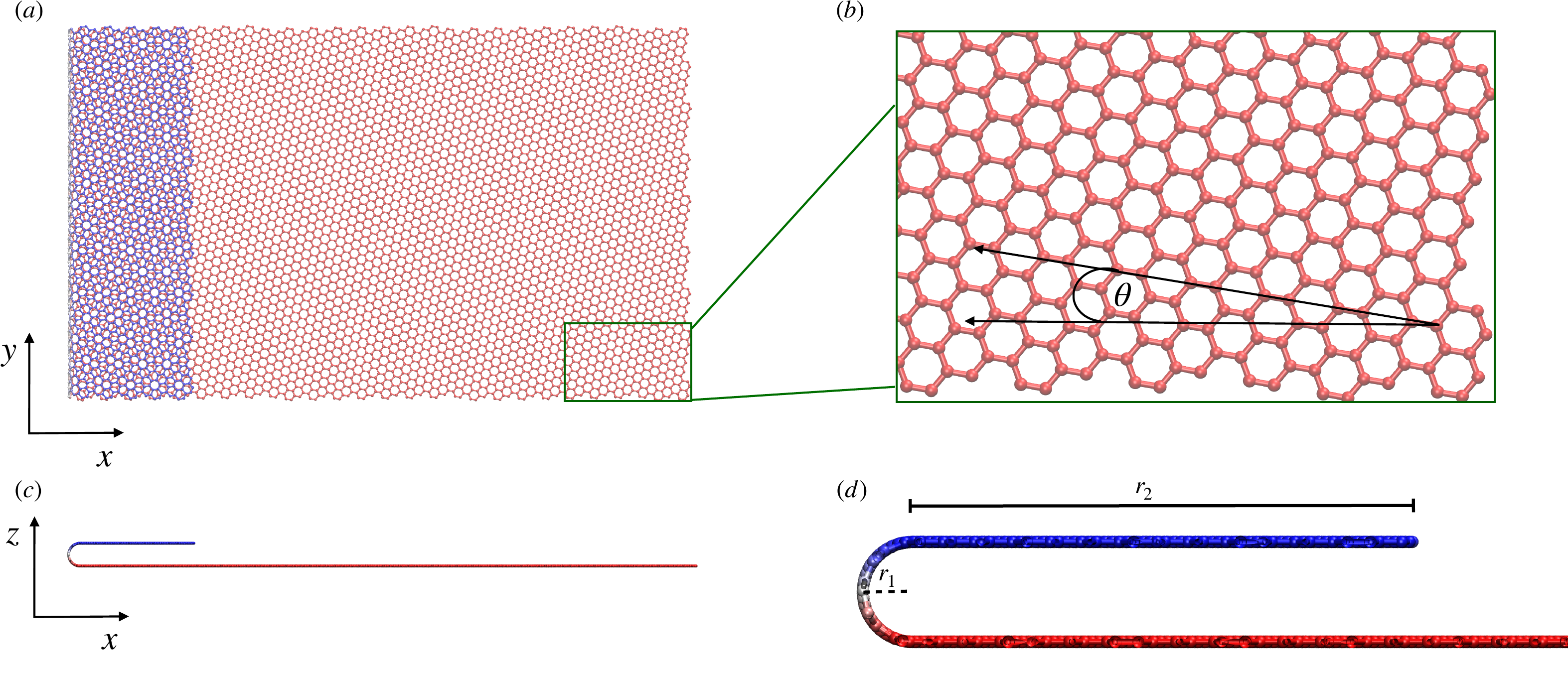}
\caption{\label{fig:layout} Schematic representation of a folded graphene sheet viewed on the $xy$ plane (a) and the $xz$ plane (c). Inserts shows the graphene lattice and how the twist angle $\theta$ (b) and the fold parameters $r_1$ and $r_2$ (d) are defined. The red and blue coloring is used to distinguish between atoms on the bottom and top layer of the fold respectively.}
\end{figure}

To simulate the growth of folded graphene structures we start simulations from folded configurations that are similar to the one shown in figure \ref{fig:layout}. MD simulations were performed both with finite sized graphene sheets and with periodic boundary conditions.  The MD parameters described in the previous section were used throughout unless otherwise stated.  Furthermore, simulations were performed both for graphene in vacuum and also for graphene on a substrate.  

\begin{figure}
    \centering
    \includegraphics[width=\linewidth]{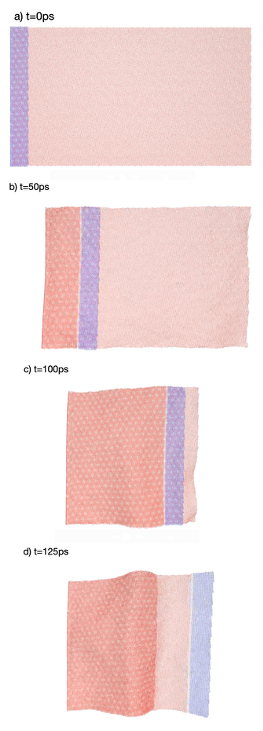}
    \caption{Example of a `successful' sliding simulation. The blue atoms are those that were initially in the top layer of the fold.  Tracking the motion of these  atoms provides a way of visualizing how the simulation progresses.}
    \label{fig:symslide}
\end{figure}

\begin{figure}
    \centering
    \includegraphics[width=\linewidth]{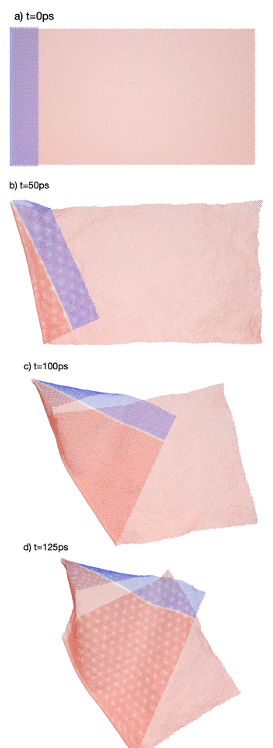}
    \caption{Same as Fig. \ref{fig:symslide} but for an `unsuccessful' sliding simulation. }
    \label{fig:asymslide}
\end{figure}

Figures \ref{fig:symslide} and \ref{fig:asymslide} show snapshots from two typical trajectories that are generated from this approach.  Figure \ref{fig:symslide} illustrates the ``ideal" behavior that is observed in these simulations.  As you can see from the figure the upper layer of the sheet slides over the bottom layer with the two edges remaining parallel throughout the simulation.  We terminate such simulations once the top layer of graphene fully occludes the bottom layer in a configuration that resembles a book.

\begin{figure}
    \centering
    \includegraphics[width=\linewidth]{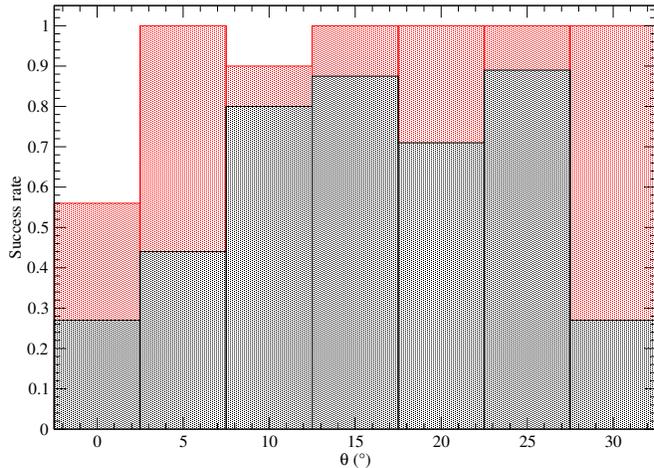}
    \caption{Success rate of simulations for various twist angles. The black and red bars are for non-periodic and periodic set-ups respectively.}
    \label{fig:error}
\end{figure}

Over half of the simulations performed behave in the manner that is illustrated in figure \ref{fig:symslide}.  However, some simulations fail either because the sliding never starts or because the sliding freezes.  Another non-ideal behavior is illustrated in figure \ref{fig:asymslide}, and involves sliding where the edges do not remain parallel throughout the simulation. All such calculations were discarded in the analysis that follows. Figure \ref{fig:error} shows the success rates for simulations with different geometries.

In a previous paper \cite{Rawlins_24} we extensively studied the formation of folded graphene structures in vacuum using a strategy similar the one that we used in this paper.  As discussed in that paper, the fold is formed parallel to the $y$-axis of the lab frame and defined by two parameters, the radius of curvature $r_1$ and the length of sheet on the top layer which overlaps with the bottom layer $r_2$.  These parameters are illustrated in figure \ref{fig:layout}. In our earlier paper we investigated what happened in the simulations for different values of these parameters. We found that folds grow if the $r_1$ parameter is set equal to 3 \AA, which is slightly greater than the typical inter layer spacing for graphite.  Furthermore, we also found that once $r_2>20$ \AA\ folding was observed consistently. We thus set $r_2$ to between values 30 and 60 \AA\ in this work.  With these parameters we were able to observe ``ideal" trajectories for both graphene in vacuum and on a substrate in all the conditions we investigated. 

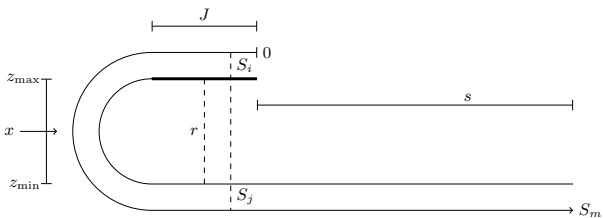
\begin{figure}
\scalebox{0.7}{\begin{tikzpicture}
\draw (8,0) -- (0,0) arc (270:90:1) -- (2,2);
\draw[<-|] (8,-0.5) node[anchor=west]{$S_m$} -- (0,-0.5) arc (270:90:1.5) -- (2,2.5) node[anchor=west]{0};
\draw[ultra thick] (2,2) -- (0,2);
\draw[dashed] (1.5,2.5) node[anchor=north west]{$S_i$}-- (1.5,-0.5) node[anchor=south west]{$S_j$};
\draw[dashed] (1, 2) -- (1, 1) node[anchor=east]{$r$} -- (1,0); 
\draw[|-|] (2,1.5) -- (8,1.5);
\node[draw=none] at (6,1.65) {$s$};
\draw[|-|] (-2,0) node[anchor=east]{$z_{\rm min}$} -- (-2,2) node[anchor=east]{$z_{\rm max}$};
\draw[|-|] (0,3) -- (1,3) node[anchor=south]{$J$} -- (2,3);
\draw[->] (-2.5,1) node[anchor=east]{$x$} -- (-1.8,1);
\end{tikzpicture}}
\caption{\label{fig:dist_diag} Schematic to illustrate how the distance $s$ between the top and bottom edge is defined in this work. The bold line on the folded sheet indicates the initial top, flat portion of the sheet. Details on how $s$ and $r$ are calculated are given in \citet{Rawlins_24}.}
\end{figure}

%Of significant importance to the following results section is the ability to measure the distance between the short edges of the sheet on the top and bottom layers.
Our earlier paper \cite{Rawlins_24} introduced a  method for calculating how the distance between the short edges of the sheet on the top and bottom layers changes as the simulation progressed.  Briefly, we assign 'safe' atoms which are always in the top and bottom layers and calculate the quantity $s_i=S_m-S_i-S_j$ for each 'safe' atom, where the $S$ parameter is the length along the folded sheet. The method that is used for this calculation is shown schematically in figure \ref{fig:dist_diag}.
%In this paper, we are only running the simulations until the top layer completely covers the bottom layer the first time, unlike in our previous paper where we observed multiple oscillations. 
%This is because in the presence of a substrate, multiple oscillations do not occur.

%A relatively simple way to examine the affect of various parameters on the sliding of graphene along itself is to time it. 
%More specifically, measure the time from the initial configuration to the first occurrence of the sheet being fully folded.
%In this paper, it is called the 'crossing time', and is shown in figure \ref{crosstimefig} for different $\theta$ values.
%In each case, the results take an average and standard error of 3-4 sample simulations for the final result and error.

In our previous paper \cite{Rawlins_24} we used a damped harmonic oscillator to model how the quantity labeled $s$ changed during our MD trajectories. This allowed us to extract a term that described the friction between the two layers and to investigate the dependence of this friction term on twist angle and temperature.  We could not use the same approach in this paper as the oscillatory motion that we observed in vacuum was not seen for graphene bound to a substrate. We thus measured the amount of friction indirectly in this work by determining a crossing time, which we define as the taken to form a fully folded structure in which $s=0$. In the next couple of sections we show that for graphene in vacuum the dependence of this quantity on the twist angle and the temperature qualitatively agrees with what we observed for the friction in our previous paper \cite{Rawlins_24}.  We then also discuss the effect of the substrate in these sections.

\section{Twist angle}
\label{sec:twist}

As discussed in section \ref{subsec:nofold} we define the twist angle ($\theta$) of the graphene as the angle between the x-axis of the lab frame and the line of symmetry that, if cut, would give rise to an edge with an 'arm-chair' configuration. Adjusting this quantity allows us to study folds in which the graphene layers are stacked commensurately as well as folds in which the graphene layers are stacked non-commensurately. Experiments \cite{Yilun_14} have shown that super lubricious sliding is only observed when the graphene layer stacking is non-commensurate. The friction between layers is considerably larger in commensurately stacked structures.  

\begin{figure}[h]
    \centering
    \includegraphics[width=\linewidth]{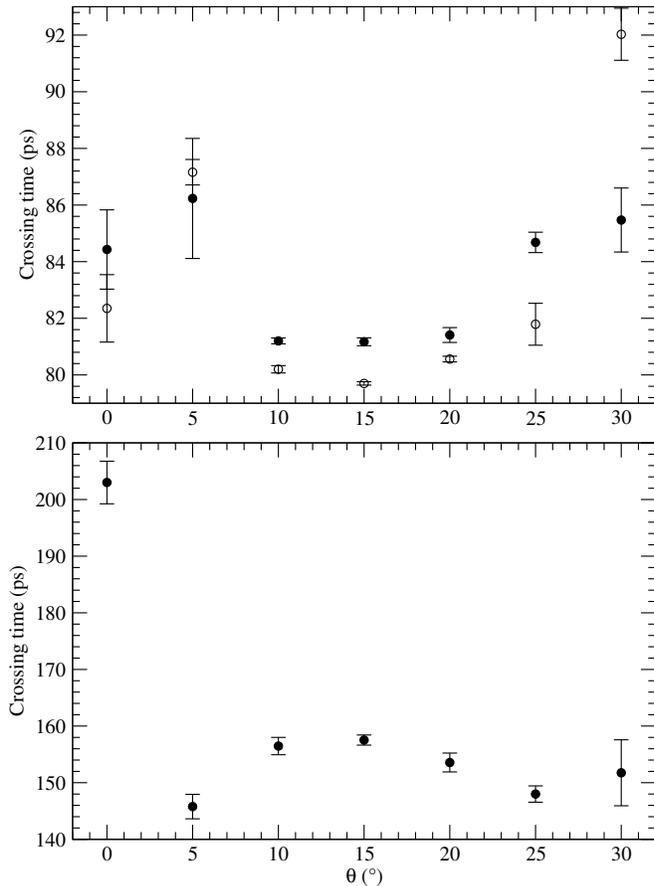}
    \caption{Crossing time for different twist angles $\theta$. The top figure contains crossing times for both non-periodic (solid black circles) and periodic (striped black circles) edges in the $y$-direction, both cases with no substrate. The bottom figure shows the crossing times for non-periodic graphene when it is supported on a substrate.}
    \label{fig:crosstime}
    \label{crosstimefig}
\end{figure}

We observed that friction was indeed higher when the graphene layers were stacked commensurately in our previous paper \cite{Rawlins_24}. Furthermore, figure \ref{fig:crosstime} indicates that the crossing times we obtain from simulations of graphene in vacuum and on a substrate behave in a manner that is consistent with what would be expected given these earlier results.  In other words, the crossing time is longer whenever the graphene sheets are stacked commensurately.

Experiments suggest that the discrepancies between the frictions in the commensurately and non-commensurately stacked structures is much larger than the differences we have observed in our simulations. Experiments have shown that the friction in these two different scenarios can differ by orders of magnitudes \cite{Dienwiebel_lubricity_04}. Although we do not see such large differences we would note that figure \ref{fig:error} shows that the majority of failed simulations were observed in simulations of commensurately stacked structures.  Consequently, the relatively low values that we observe for the crossing time of these commensurately-stacked structures is likely affected by the trajectories we have analyzed. If we had the computational time to run on the simulations where the growth stopped they may have formed a fully folded structure over the course of a much longer simulation time.  In other words, slow dynamics is observed for commensurately stacked structures because commensurately stacked structures are more likely to undergo a stick slip motion. We have to deliberately exclude sticking in our simulations to reduce computational expense and are thus investigating differences between dynamic frictions in the commensurate and non-commensurate regimes, which are likely smaller.      

A comparison of the top and bottom panels of figure \ref{fig:crosstime} suggest that when the substrate is introduced there is more resistance to fold growth. In general the crossing time for both the commensurately and non-commensurately stacked structures increases by a factor of two. 

\section{Temperature}
\label{sec:temperature}

\begin{figure}
    \centering
    \includegraphics[scale=0.4]{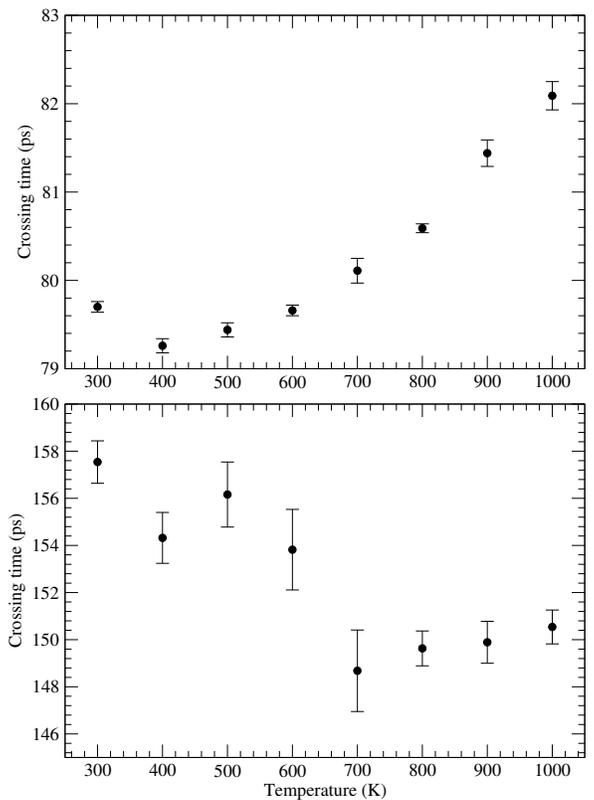}
    \caption{Crossing times for sliding simulations with different temperatures. Top (bottom) figure corresponds to sliding simulations without (with) a substrate. All simulations were performed with $\theta=15^\circ$ and with periodic boundary conditions.}
    \label{fig:Temptime}
\end{figure}

To investigate the effect that temperature has on the growth of these folded graphene sheets we ran multiple simulations starting from a folded graphene sheet with a twist angle of 15$^\circ$.  All these calculations were performed with periodic boundary conditions as simulations of finite-sized graphene sheets at high temperature are very likely to fail in the manner illustrated in figure \ref{fig:asymslide}.  The crossing times as a function of temperature that we obtained from simulations of the graphene in vacuum and on a substrate are shown in figure \ref{fig:Temptime}.

In our previous paper \cite{Rawlins_24} we found that friction increased with temperature for graphene in vacuum, which is consistent with the results shown in the upper panel of figure \ref{fig:Temptime}. This figure clearly shows that the crossing time increases with temperature.  In our paper \cite{Rawlins_24} we argued that this increase occurs because increases in temperature cause the graphene sheet to undergo greater out-of-plane oscillations.  These increased fluctuations make the surface appear rougher and thus more strongly suppress sliding motions. 

The behavior we observe when the graphene is on a substrate is consistent with this interpretation of the behavior in vacuum.  As the lower panel of figure \ref{fig:Temptime} shows, when supported graphene is simulated the crossing time no longer increases with temperature. This increase is no longer observed because the presence of the substrate restricts the out of plane motions that cause the slow downs that are observed for graphene in vacuum.

\section{Extracting interfacial free energies from simulations of sliding}
\label{sec:extract}

It is straightforward to show that the quantity $s$ that is monitored in our simulations and that is illustrated in figure \ref{fig:dist_diag} is related to $c$ in equation \ref{eqn:driving} by:
$$
s=l_x-2c
$$
where $l_x$ is the total length of the graphene sheet.  Inserting this result into equation \ref{eqn:driving} and setting $2\lambda c = 0$, because there is no tearing in our simulations, gives:
%In our simulations, there is no tearing so the second term is set to zero.
%If we assume that the area of the fold is negligible compared to the total area of the sheet, then $A_{ribbon}$ can be written in terms of the distance between the short edges of the sheet $d$, $A_{ribbon}=l_y(l_x-d)/2$.
%Therefore, Eq. \ref{Eq_energy} can be rewritten as:
\begin{equation}
    U=U_{fold}-\frac{l_xw}{2}(\gamma_{gg}+\gamma_{gs})+\frac{w}{2}(\gamma_{gg}-\gamma_{gs})s.
    \label{Equpdate_energy}
\end{equation}
This result suggests that we can extract $U_{fold}, \gamma_{gg}$ and $\gamma_{gs}$ by plotting the $U$ values from our simulations against $s$ values and doing a linear regression.

\begin{figure}
    \centering
    \includegraphics[width=\linewidth]{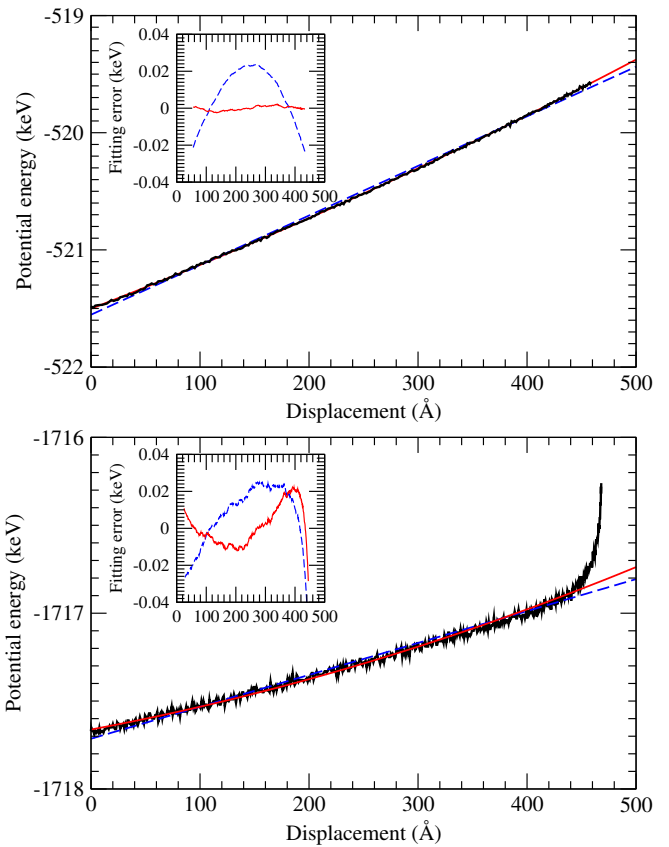}
    \caption{Example of fitting the potential energy as a function of displacement for $\theta=15^\circ$. The top panel shows the result for the simulation without the substrate, while the bottom panel is the result from the simulation with a substrate. In both panels the black lines show the raw simulation data. The result from fitting this data with the linear model described in the text is shown using a blue dashed line while the solid red line is the fit that is obtained with the quadratic model. The fact that the quadratic model fits the data most closely is best illustrated by the insets, which show the fitting error as a function of the displacement.
    %Top (bottom) figures show the raw (black lines) and fitting formula (red line) for simulations without (with) a substrate.
    }
    \label{fig:Efit}
\end{figure}

The black lines in figure \ref{fig:Efit} illustrates that the results from our simulation are consistent with the prediction of this model. This figure shows how the potential energy changes with $s$ in simulations of a fold with $\theta=15^\circ$. When the graphene is in vacuum (top panel) and the graphene is on a substrate (bottom panel) the potential energy does appear to be a linear function of $d$.  When the substrate is present, however, there is some deviation from this linear behavior at the start of the simulation, when $d$ is large, because it takes longer for the graphene sheet to equilibrate when it is on a substrate.   

The blue dashed lines in figure \ref{fig:Efit} show the the result that is obtained when equation \ref{Equpdate_energy} is used to fit the $E$ versus $s$ data for all $s$ values that are less than 400 \AA.  If you look at the inset for the top panel you can see that the data is not uniformly scattered around this line. The model appears to consistently overestimate the energy for values of $s$ between 150\AA\ and 250\AA\ and underestimate it when $s$ is outside this range. Furthermore, fitting the simulation without a substrate gives $\gamma_{gg}=0.449 \pm 0.002$ Jm$^{-2}$, which is considerably larger than the values for $\gamma_{gg}$ that we obtained in section \ref{subsec:nofold}. We thus introduced an additional quadratic term and refit the data using:
\begin{equation}
    U=U_{fold}-\frac{l_xw}{2}(\gamma_{gg}+\gamma_{gs})+\frac{w}{2}(\gamma_{gg}-\gamma_{gs})s+\frac{k}{2}s^2.
    \label{Eq_Ek2}
\end{equation}
The results obtained with this model are shown in red in figure \ref{fig:Efit}. Introducing this quadratic term gives an improved fit to the data and also lowers the estimate of $\gamma_{gg}$ to $0.378\pm 0.004$ Jm$^{-2}$, which is still larger than the values we obtained in section \ref{subsec:nofold}.

An alternative approach for extracting the adhesion parameters involves looking at the forces.
Taking the negative derivative of Eq. \ref{Eq_Ek2} with respect to $s$ and introducing $\Delta\gamma=\gamma_{gg}-\gamma_{gs}$ gives:
\begin{equation}
    -\frac{dU}{ds(t)}=-\frac{w}{2}\Delta\gamma-\kappa s(t) = m\frac{d^2s(t)}{dt^2}.
\end{equation}
In the final equality of the equation above we relate the force that is obtained by taking the negative derivative with the acceleration by introducing $m$, the mass of the ribbon, which we have assumed is a constant.
We have also replaced $k$ with $\kappa$ to show that while this term originates in equation \ref{Eq_Ek2} it may not behave the same when using a new fitting procedure.
%Since the only part of the sheet that is moving is the top layer, it would be more accurate to have $m$ be the mass of the top layer as a function of $d(t)$.
%But having a variable mass makes the overall differential equation unwieldy, so instead we use a constant mass.
To account for friction due to sliding, we add a damping term to this differential equation. 
After dividing through by $m$ we arrive at:
\begin{equation}
    \frac{d^2s(t)}{dt^2}=-\frac{w}{2m}\Delta\gamma-\frac{\kappa}{m}s(t)-\frac{b}{m}\frac{ds(t)}{dt},
    \label{Eq_diff}
\end{equation}
where $b$ is a kinetic drag coefficient.

%\begin{figure}
%\scalebox{0.9}{\begin{tikzpicture}
%\draw (9,0) -- (0,0) arc (270:90:1) -- (3,2);
%\draw[ultra thick] (3,2) -- (1,2);
%\draw[|-|] (3,1.75) -- (9,1.75);
%\node[draw=none] at (7,1.9) {$d(t)$};
%\draw[|-|] (1,2.5) -- (2,2.5) node[anchor=south]{$r_2$} -- (3,2.5);
%\draw[dashed] (2,2) -- (2,0);
%\draw[->] (2,0.5) -- (3,0.5) node[anchor=west]{$F_{gg}$};
%\draw[->] (2,0.4) -- (1,0.4) node[anchor=east]{$F_{gs}$};
%\draw[->] (2,0.9) -- (1,0.9) node[anchor=east]{$c\frac{\delta d(t)}{\delta t}$};
%\draw[->] (2,1.4) -- (1,1.4) node[anchor=east]{$kd(t)$};
%\end{tikzpicture}}
%\caption{Force diagram for a sliding folded sheet of graphene. }
%\label{fig:forcediag}
%\end{figure}

The solution of this differential equation is,
\begin{equation}
    \begin{split}
        s(t)&=-\frac{w}{2\kappa}\Delta\gamma+e^{-\frac{t}{2}\frac{b}{m}}\left(S_0+\frac{w}{2\kappa}\Delta\gamma\right)\\
        &\times\left(\frac{b/m}{\sqrt{\left(\frac{b}{m}\right)^2-\frac{4\kappa}{m}}}\sinh\left(\frac{t}{2}\sqrt{\left(\frac{b}{m}\right)^2-\frac{4\kappa}{m}}\right)\right.\\
        &\left.+\cosh\left(\frac{t}{2}\sqrt{\left(\frac{b}{m}\right)^2-\frac{4\kappa}{m}}\right)\right)
    \end{split}
    \label{Eq_dist}
\end{equation}
where $S_0$ is some initial displacement.

The velocity can also be expressed as,
\begin{equation}
    \begin{split}
        \frac{ds(t)}{dt} &=-\frac{2\kappa}{m}e^{-\frac{t}{2}\frac{b}{m}}\left(S_0+\frac{w}{2\kappa}\Delta\gamma\right) \\
        &\times\frac{\sinh\left(\frac{t}{2}\sqrt{\left(\frac{b}{m}\right)^2-\frac{4\kappa}{m}}\right)}{\sqrt{\left(\frac{b}{m}\right)^2-\frac{4\kappa}{m}}}
    \end{split}
    \label{Eq_vel}
    \end{equation}

Performing the same procedure with equation \ref{Equpdate_energy}, which sets $\kappa=0$ and thus neglects the harmonic term, results in the following expressions:
\begin{equation}
    \begin{split}
        s(t)&=S_0-\frac{w\Delta\gamma}{2b}\left(t-\frac{m}{b}\left(1-e^{-t\frac{b}{m}}\right)\right)\\
        \frac{ds(t)}{dt}(t)&= -\frac{w\Delta\gamma}{2b}\left(1-e^{-t\frac{b}{m}}\right)
    \end{split}
    \label{Eq_k0}
\end{equation}

\begin{figure}
    \centering
    \includegraphics[width=\linewidth]{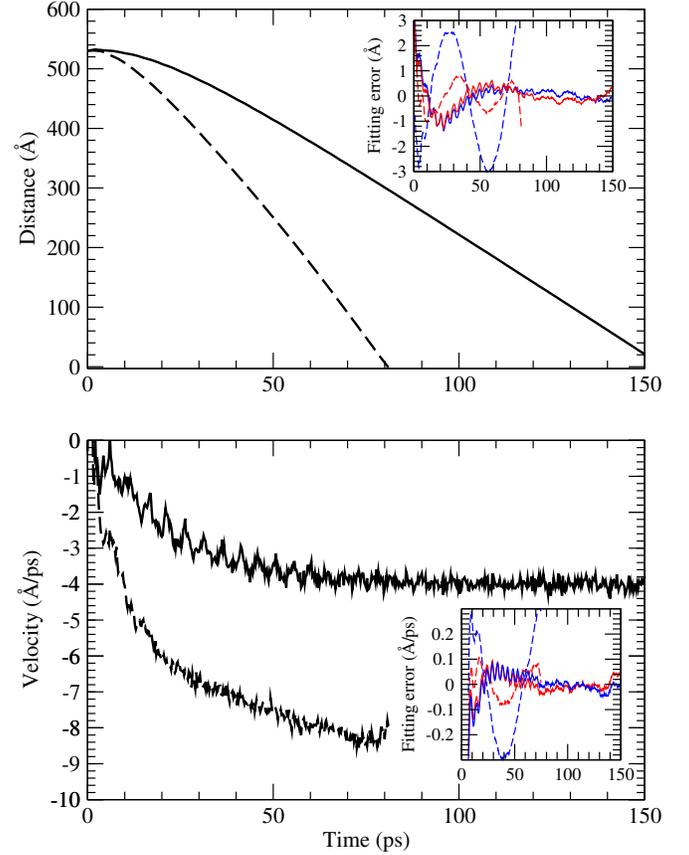}
    \caption{Example fitting results for displacement (top figure) and velocity (bottom figure) versus time curves from simulations with $\theta=15^\circ$. The main graphs show the raw data displacement and velocity data from the MD simulations without a substrate (dashed line) and without a substrate (solid line).  Notice that when the substrate is present it takes longer for the system to arrive to the fully folded configuration and that the fold velocity has a lower magnitude. The insets show the errors as a function of simulation time for fits of this data to equation \ref{Eq_k0} (blue) and to equations \ref{Eq_dist} and \ref{Eq_vel} (red) with dashed and solid lines being used for simulations with and without a substrate as in the main figures.
    %Black (blue) lines are the raw data for the displacement and velocity without (with) a substrate, and the red lines are the fitting functions for displacement (Eq. \ref{Eq_dist}) and velocity (Eq. \ref{Eq_vel}).
    }
    \label{fig:forcefit}
\end{figure}

Figure \ref{fig:forcefit} shows displacement and velocity versus time curves from simulations of graphene in vacuum and graphene on a substrate with $\theta=15^\circ$. The insets in these figures show results that are obtained when these curves are fit using equations \ref{Eq_dist}, \ref{Eq_vel} and \ref{Eq_k0}. The fitting error as a function of simulation time is displayed in these insets.  These results indicate that when the graphene is simulated in vacuum equations \ref{Eq_dist} and \ref{Eq_vel} provide a closer fit to the simulation data than equations \ref{Eq_k0}. By contrast, when simulations of graphene on the substrate are fitted the two models describe the data equally well. Further evidence that the harmonic terms are only important when graphene is in vacuum is provided by the fitted values for $\kappa$, which are one to two orders of magnitude smaller when the fitted simulation is of graphene on a substrate. The change in the fitted value for $\Delta \gamma$ when moving from equation \ref{Eq_k0} to equations \ref{Eq_dist} and \ref{Eq_vel} is also smaller when the fitted simulation is of graphene supported on a substrate.

The physical origin of the harmonic terms in equations \ref{Eq_dist} and \ref{Eq_vel} is unclear.  The fact that the fits reported in figure \ref{fig:forcefit} show that this term is more important in simulations of graphene in vacuum perhaps suggests that this term is capturing the effect of out-of-plane oscillations in the graphene that are not present when the graphene is supported on a substrate.  However, we would note that when the potential energy data from simulations of graphene on a substrate are fitted using equations \ref{Equpdate_energy} and \ref{Eq_Ek2} we obtain $\Delta \gamma$ values of 0.188 and 0.121~J/m$^{2}$. Including or not including $k$ when fitting this data on supported graphene thus clearly has a significant effect.  Martini and co workers suggest that higher order terms are required to describe the changes in potential that occur as the lattice offset between the top and bottom layers changes during sliding.  This factor, unlike the out of plane motions, is equally likely to affect graphene in vacuum and graphene on a substrate.  Our results perhaps suggest that this phenomenon is seen when the energy is plotted as a function of $s$.  However, these subtle variations in energy due to lattice offset have little effect on $s(t)$ and $\textrm{d} s(t) / \textrm{d}t$.  In other words, the physical origin of the harmonic terms in equation \ref{Eq_Ek2} may be different from the origins of this term in equations \ref{Eq_dist} and \ref{Eq_vel}.  

%Looking at figure \ref{fig:forcefit}, both the non-substrate and substrate displacement plots are very smooth.
%The both of the fitting formulas Eq. \ref{Eq_dist} and \ref{Eq_k0} fit the data very well at first glance, but if we examine the error plot it can be seen that by including the harmonic term in the non-substrate fitting results in a noticeable drop in the error. 
%This is reflected in the calculated $\gamma_{gg}$ from this particular example.
%With $k=0$, $\gamma_{gg}=0.268\pm0.013$ and $0.271\pm0.013$ J/m$^2$ and $\Delta\gamma=0.087\pm0.003$ and $0.087\pm0.003$ J/m$^2$ from the displacement and velocity fits respectively.
%Including $k$, $\gamma_{gg}=0.381\pm0.006$ and $0.384\pm0.006$ J/m$^2$ and $\Delta\gamma=0.090\pm0.002$ and $0.076\pm0.001$ J/m$^2$ from the displacement and velocity fits respectively.
%If $k$ is included in the fit, the calculated $\gamma_{gg}$ value is 40\% lower than if $k$ is not included.
%The same cannot be said for the substrate calculations, where the inclusion of $k$ makes no noticeable difference.

\begin{figure}
    \centering
    \includegraphics[width=\linewidth]{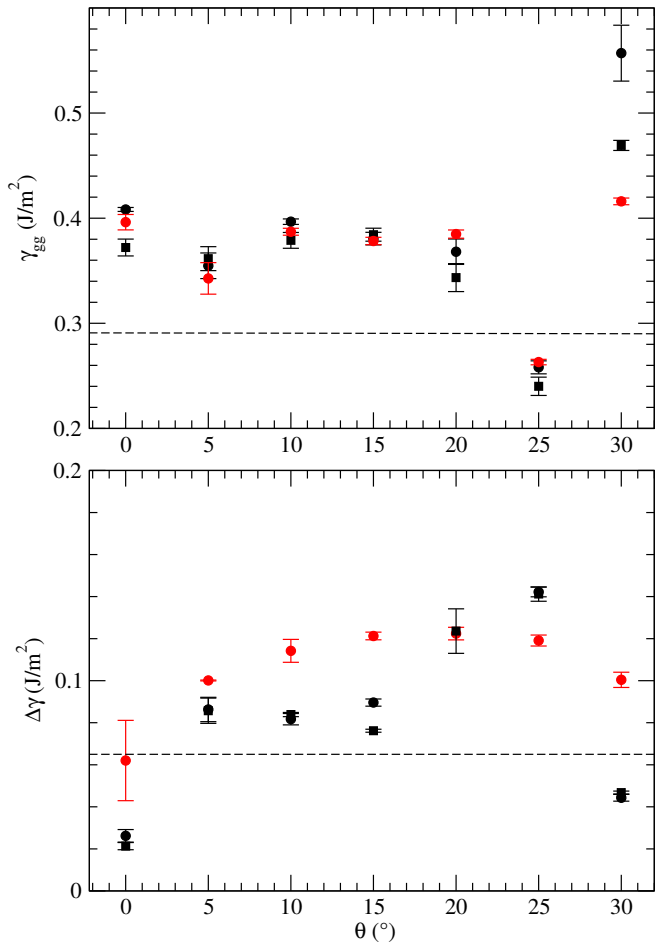}
    \caption{Calculated adhesion values determined by fitting the trajectories discussed in section \ref{sec:twist} using equations \ref{Eq_Ek2} (red circles), \ref{Eq_dist} (black circles) and \ref{Eq_vel} (black squares). The top panel shows $\gamma_{gg}$ values that were obtained from simulations of graphene in vacuum, while the bottom panel shows $\Delta\gamma$ that were obtained from simulations of graphene on a substrate.} 
    \label{fig:gamma}
\end{figure}

To test the models for sliding described above we used them to fit the trajectories that were discussed in sections \ref{sec:twist} and \ref{sec:temperature}. Figure \ref{fig:gamma} shows the $\gamma_{gg}$ and $\Delta \gamma$ values that were obtained from the trajectories that were introduced in section \ref{sec:twist}. Similar results that were obtained when the data from section \ref{sec:temperature} were fitted are provided in the supporting information.

Figure \ref{fig:gamma} shows that the estimates obtained for $\gamma_{gg}$ and $\Delta \gamma$ by fitting trajectories using equations \ref{Eq_Ek2}, \ref{Eq_dist} and \ref{Eq_vel} are mostly consistent with each other.  Furthermore, figure \ref{fig:gamma},  like figure \ref{fig:layeradhesion}, shows that the adhesion energies do not have a particularly strong dependence on twist angle. In other words, when you change $\theta$, $\gamma_{gg}$ and $\Delta \gamma$ do not change. The differences in crossing times that were observed in figure \ref{fig:crosstime} do not occur because $\gamma_{gg}$ and $\Delta \gamma$ have different values for different graphene geometries.

It is interesting to note that the $\gamma_{gg}$ and $\Delta \gamma$ values that are obtained by fitting are almost always higher than the values of these quantities that were obtained through the static calculations. We believe that this difference occurs because the growth is an irreversible process. As the fold grows in our simulations energy is not simply transferred between the graphene-substrate and the graphene-graphene interactions.  Energy is also transferred to the environment and these additional transfers of energy ensure that the $\Delta \gamma$ values that emerge from fitting are larger than the $\Delta \gamma$ values that are obtained from static calculations.

\section{Substrate adhesion energy\label{subsec:LJ}}

As discussed in the previous section the crossing time is not only a function of $\Delta \gamma$. The crossing time strongly depends on the twist angle, which, as figures \ref{fig:layeradhesion} and \ref{fig:gamma} show, has little effect on $\Delta \gamma$. However, it is not correct to state that the crossing time does not depend on $\Delta \gamma$.  Crossing times for graphene in vacuum are shorter than those for graphene on a substrate because $\gamma_{gg}>\gamma_{gg}-\gamma_{gs}$. 

\begin{figure}
    \centering
    \includegraphics[width=\linewidth]{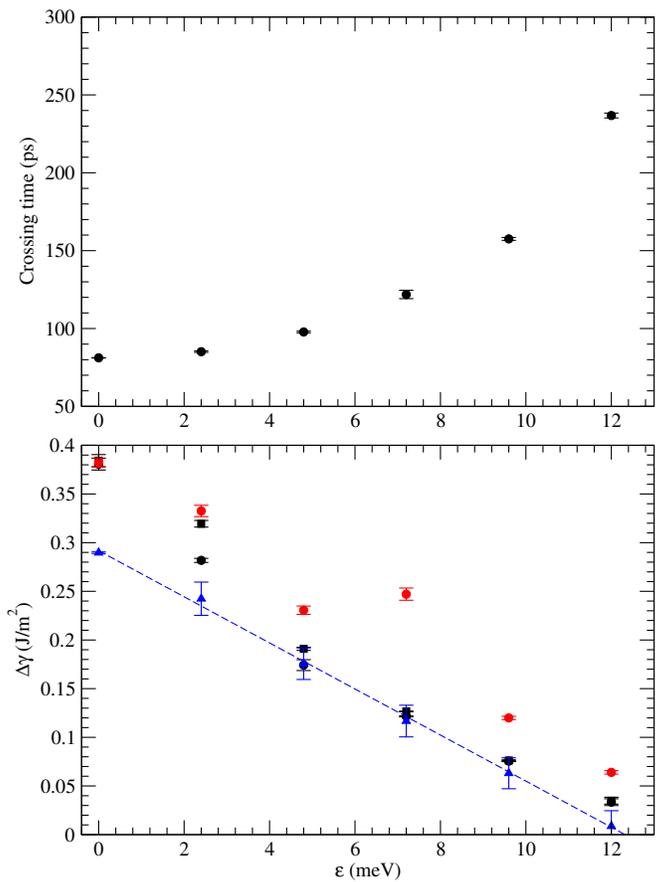}
    \caption{Crossing time (top figure) and difference in adhesion energies $\Delta\gamma$ (bottom figure) for simulations of graphene on various substrates. To model the effect of different substrates we used a range of values for the $\epsilon$ parameter in the Lennard-Jones potential that describes the interaction between the carbon atoms and the atoms in the substrate. $\Delta \gamma$ values were obtained by fitting the trajectories using equations \ref{Eq_Ek2} (red), \ref{Eq_dist} (black circles) and \ref{Eq_vel} (black squares). For comparison we also calculated $\Delta \gamma$ values using static calculations similar to those described in section \ref{subsec:nofold}. The values obtained from these calculations are indicated with blue triangles.}
    \label{fig:LG}
\end{figure}

To investigate whether there is a simple relationship between the $\Delta \gamma$ value and the crossing times we performed simulations of a folded graphene sheet with a twist angle of 15$^\circ$ on a range of different substrates. To model substrates with different strengths of binding to graphene we adjusted the $\epsilon$ parameter of the Lennard Jones interaction for the substrate graphene potential.  The results obtained from these simulations are shown in figure \ref{fig:LG}.

The top panel of figure \ref{fig:LG} shows that the crossing time increases monotonically as you increase the $\epsilon$ parameter and make the substrate bind more strongly to the graphene.  Not shown in figure \ref{fig:LG} is the fact that in simulations with $\epsilon>14$ the graphene sheet always unfolds so the crossing time becomes essentially infinite when $\epsilon$ is large.

The bottom panel of figure \ref{fig:LG} shows that there are close to linear decreases in the value of $\Delta \gamma$ as $\epsilon$ increases regardless of whether this quantity is computed statically or by fitting the trajectories using equations \ref{Eq_Ek2}, \ref{Eq_dist} and \ref{Eq_vel}. The blue dashed line in figure \ref{fig:LG} shows a linear fit for the results from the static calculations and suggests that $\Delta \gamma$ is negative for all $\epsilon>12.3$~meV. This result is consistent with our observation that graphene unfolds in simulations with $\epsilon=14$~meV.

As was observed in the previous section, the $\Delta \gamma$ values that emerge from fitting are always larger than the values of this quantity that are obtained from static calculations.  Interestingly, the estimates for $\Delta \gamma$ from equations \ref{Eq_Ek2}, \ref{Eq_dist} and \ref{Eq_vel} are almost identical when $\epsilon=0$ and the substrate is absent.  Once the substrate is introduced there are discrepancies between the three estimates.  This perhaps suggests that the greater conformational flexibility of graphene in vacuum cancels out terms that are missing from the three models. When modelling graphene on a substrate models that accounting for these additional terms that perhaps describe how the mass of the fold change as it grows is necessary.

%Adjusting the strength of the LJ potential between the graphene sheet and substrate allows for insight into the interplay between the graphene-graphene adhesion and graphene-substrate adhesion.
%Figure \ref{fig:LG} shows the behaviour of crossing time and $\Delta\gamma$ as a function of $\sigma$.
%The figures readily shows that as $\sigma$ is increased, the crossing time is increased and $\Delta\gamma$ decreases. For $\Delta\gamma$, all three adhesion values decrease with $\sigma$, however, the actual values differ between methods.

%\subsection{Temperature\label{subsec:temp}}

% \begin{figure}
%     \centering
%     \includegraphics[scale=0.4]{Temp_gamma.eps}
%     \caption{Same as Fig. \ref{fig:gamma}, except for different temperature values. All simulations were performed with $\theta=15$ and with periodic boundary conditions.}
%     \label{fig:Tempgamma}
% \end{figure}
\section{Conclusions}

As discussed in the introduction the difference between  the graphene-substrate and graphene-graphene adhesion energies is the thermodynamic driving force for autokirigami formation and the formation of folded graphene structures in general. In this work we have discussed four methods for extracting these two quantities from simulations. The first is the standard static approach of calculating surface energies.  The remaining three involve running MD simulations of folded graphene sheets and fitting the data using an energy balance model. The $\gamma_{gg}$ and $\Delta\gamma$ values we extract using these four different methods do not have a strong dependence on twist angle or temperature.  However, the adhesion energies that emerge from fitting are almost always larger than the values obtained from static calculations.  We attribute this result to the fact that our simulations are irreversible, while the energy balance model assumes a reversible dynamics.

The thermodynamic driving force that causes graphene folds to grow is given by $\Delta \gamma = \gamma_{gg} - \gamma_{gs}$. Our results show that the magnitude of this difference has an effect on the speed at which the folded structures forms but also that this is not the only factor that determines the kinetics.  Changing the value of the  twist angle, $\theta$, changes for the speed of fold growth significantly.  However, the values for $\gamma_{gg}$ and $\gamma_{gs}$ do not depend on $\theta$.  Clearly, something more than an energy balance model is required for understanding the kinetics of autokirigami growth.  

\begin{acknowledgments}
This work was supported by a research grant from the Department for the Economy Northern Ireland under the US-Ireland R\&D Partnership Programme. The computing resources used for this publication were from the Northern Ireland High Performance Computing (NI-HPC) service funded by EPSRC (EP/T022175).
\end{acknowledgments}

\section*{Data Availability Statement}
The data that supports the findings of this study are available from the corresponding author upon reasonable request. 
Scripts for generating folded graphene sheets, input files for running on LAMMPS and scripts used to extract data from the resulting simulations as described in section \ref{sec:method} will be available on \href{https://github.com/crawlins5/Graphene_folding.git}{Github}.

%\nocite{*}
\bibliography{references}% Produces the bibliography via BibTeX.

\end{document}